\documentclass{PoS}

\usepackage{amsmath,amssymb,cite,graphicx}

\title{On the light massive flavor dependence of the top quark mass}

\ShortTitle{On the light massive flavor dependence of the top quark mass}

\author{Andr\'e H. Hoang\\
        University of Vienna, Vienna, AUSTRIA\\
        E-mail: \email{andre.hoang@univie.ac.at}}

\author{\speaker{Christopher Lepenik}\\
        University of Vienna, Vienna, AUSTRIA\\
        E-mail: \email{christopher.lepenik@univie.ac.at}}
        
\author{Moritz Preisser\\
        University of Vienna, Vienna, AUSTRIA\\
        E-mail: \email{moritz.preisser@univie.ac.at}}
        
\newcommand{\nl}{n_{\ell}}
\newcommand{\dd}{\mathrm{d}}
\newcommand{\MSb}{\overline{\mathrm{MS}}}
\newcommand{\mbar}{\overline{m}}
\newcommand{\mpole}{m^{\mathrm{pole}}}
\newcommand{\Ord}{\mathcal{O}}
\newcommand{\LQCD}{\Lambda_\mathrm{QCD}}
\newcommand{\delbar}{\overline{\delta}}
\newcommand{\nq}{n_Q}
\newcommand{\msr}{m^{\rm MSR}}
\newcommand{\MSR}{\mathrm{MSR}}
\newcommand{\pole}{\mathrm{pole}}
\newcommand{\betazero}{\beta_0^{(\nq)}}
\newcommand{\betaone}{\beta_1^{(\nq)}}

\abstract{We provide a systematic renormalization group formalism to study the mass effects in the relation of the pole mass and short-distance masses such as the $\MSb$ mass of a heavy quark $Q$, coming from virtual loop insertions of massive quarks lighter than $Q$ with the main focus on the top quark.
The formalism reflects the constraints from heavy quark symmetry and entails a combined matching and evolution procedure that allows to disentangle and successively integrate out the corrections coming from the lighter massive quarks and the momentum regions between them and also to precisely control the large order asymptotic behavior.
The formalism is used to study the asymptotic behavior of light massive flavor contributions and is applied to predict the ${\cal O}(\alpha_s^4)$ virtual quark mass corrections, calculate the pole mass differences for massive quark flavors with a precision of around $20$~MeV, and determine the pole mass ambiguity which amounts to $250$~MeV in the physical case of three massless quark flavors.\\[.5cm]
UWThPh-2018-6
}

\FullConference{13th International Symposium on Radiative Corrections (Applications of Quantum Field Theory to Phenomenology)\\
                 25-29 September, 2017 \\
                 St. Gilgen, Austria}

\begin{document}

\section{Introduction}\label{sec:intro}

The masses of the heavy quarks, especially of the top, are among the most important parameters in the standard model, having strong impact on e.g.\ precise consistency tests of the standard model and the estimation of the electroweak vacuum stability. However, it must be kept in mind that the mass of a heavy quark $Q$, due to confinement, is not a physical observable but should be viewed as a formal theory parameter which depends on the renormalization scheme. 
Depending on the observable and energy scale of interest different renormalization schemes are used in calculations to minimize theoretical uncertainties. Therefore it is necessary to determine precise relations between the several mass renormalization schemes such that accurate comparisons can be made between them.

Considering the pole mass scheme, it is well known that the resulting mass parameter $\mpole_Q$ is linearly sensitive to small momenta and hence sensitive to the non-perturbative regime of QCD. This low momentum sensitivity even grows rapidly with loop order and leads to the so-called $\Ord(\LQCD)$ renormalon \cite{Bigi:1994em,Beneke:1994sw,Beneke:1998ui}. On the other hand there are ``short-distance masses'' like the $\MSb$, kinetic~\cite{Czarnecki:1997sz}, PS~\cite{Beneke:1998rk}, 1S~\cite{Hoang:1998ng,Hoang:1998hm,Hoang:1999ye}, RS~\cite{Pineda:2001zq}, and MSR~\cite{Hoang:2008yj,Hoang:2017suc} masses which have no linear low momentum sensitivity and consequently do not have such a renormalon.
The $\MSb$ mass $\mbar_Q(\mu)$ is defined analogous to the $\MSb$-renormalized strong coupling constant $\alpha_s(\mu)$. Like the strong coupling, the $\MSb$ mass $\mbar_Q(\mu)$ depends on a renormalization scale $\mu$
which should be parametrically of the order or or higher than the mass scale itself.
In the case of the $\MSb$ mass this scale can be interpreted as the scale above which short-distance information from on-shell self-energy diagrams is contained in the mass. So the difference between $\MSb$ and pole mass, $\mpole-\mbar_Q(\mu)$, contains these self-energy contributions between momentum zero and the scale $\mu$.

In the approximation that all flavors lighter than the heavy quark $Q$ are massless the relation between the pole and $\MSb$ mass can be written in the form
\begin{equation}
\mpole_Q - \mbar_Q = \mbar_Q\,\sum_{n=1}^\infty\,a_n(\nq,n_h)\,\left(\frac{\alpha_s^{(\nq+n_h)}(\mbar_Q)}{4\pi}\right)^n,
\end{equation}
where $\mbar_Q\equiv\mbar_Q(\mbar_Q)$, $n_Q$ is the number of quark flavors lighter than $\mbar_Q$ (which are taken massless at this point), $n_h$ is the number of quark flavors with mass $\mbar_Q$, and $\alpha_s^{(\nq+n_h)}$ is the strong coupling constant that evolves with $\nq+n_h$ active dynamical flavors according to the evolution equation
\begin{equation}\label{eqn:betafct}
\frac{\dd\alpha_s^{(n_Q)}(\mu)}{\dd\log \mu}=\beta^{\,(n_Q)}(\alpha_s(\mu))\,=\,-\,2\,\alpha_s^{(n_Q)}(\mu)\sum_{n=0}^\infty\beta_n^{\,(n_Q)}\bigg(\frac{\alpha_s^{(n_Q)}(\mu)}{4\pi}\bigg)^{n+1}\,.
\end{equation}

The perturbative coefficients $a_n(\nq,n_h)$ are known up to $\Ord(\alpha_s^4)$ from explicit loop calculations~\cite{Tarrach:1980up, Gray:1990yh,Chetyrkin:1999ys,Chetyrkin:1999qi,Melnikov:2000qh,Marquard:2007uj,Marquard:2015qpa,Marquard:2016dcn}. Owing to their renormalon behavior, they are known asymptotically to all orders through formulas like~\cite{Beneke:1998ui,Hoang:2017suc}
\begin{equation}
\label{eqn:anintermsofsk}
a_n^\mathrm{asy}(n_Q,n_h)= a_n^\mathrm{asy}(n_Q,0)=4\pi N_{1/2}^{(n_Q)}(2\beta_0^{(n_Q)})^{n-1}\sum_{k=0}^\infty g_k^{(n_Q)}\frac{\Gamma(n+\hat b_1^{(n_Q)}-k)}{\Gamma(1+\hat b_1^{(n_Q)})}\,,
\end{equation}
where $g_\ell$ and $\hat b_1$ are polynomials of the QCD $\beta$-function coefficients $\beta_n$ (see Eq.~\eqref{eqn:betafct}) and the anomalous dimension of the $\MSb$ mass~\cite{Hoang:2017suc}, and $N_{1/2}^{(n_Q)}$ is a normalization~\cite{Ayala:2014yxa,Beneke:2016cbu,Hoang:2017suc}. Numerically, the divergence pattern depends strongly on the massless flavor number $n_Q$.
It is intriguing that already the 4-loop coefficient follows the asymptotic behavior quite closely:
\begin{align}
a^{\mathrm{asy}}_4(n_Q=4,1)&=230192\pm 14747,\quad\text{\cite{Hoang:2017suc}}\\
a_4(n_Q=4,1)&=214828\pm 422.\hphantom{47}\hspace{1pt}\quad\text{\cite{Marquard:2016dcn}}
\end{align}

\section{Bottom and Charm Mass Dependence}\label{sec:MassDependence}

Due to the hierarchy in quark masses, in many applications of heavy quark physics lighter massive quarks may be taken as massless. However,
since the pole mass is linearly sensitive to low momenta, it is sensitive to lighter massive quark flavors $q$ ($m_q>\LQCD$) as well. The impact is relevant especially at high orders:
the mass of a virtual quark flavor in an on-shell self-energy diagram acts as an effective infrared cut-off at the mass scale and therefore this quark flavor effectively decouples at high orders where the series is governed by scales $\lesssim\LQCD$~\cite{Ball:1995ni}.
It is obvious that, since the masses of the lighter massive flavors alter the pattern of divergence of the renormalon series, their induced corrections are themselves plagued by a renormalon.

The effects of massive lighter flavors in the pole-$\MSb$ mass relation are known through explicit loop calculation up to $\Ord(\alpha_s^3)$~\cite{Gray:1990yh,Bekavac:2007tk} and, as expected, these corrections are not convergent due to the contained renormalon. Prior to our work~\cite{Hoang:2017btd} the large-order asymptotic behavior of these corrections and a systematic approach to the flavor decoupling described in the previous paragraph was unknown. In the this talk we discuss Ref.~\cite{Hoang:2017btd} where we introduced a renormalization group framework, which is capable of describing exactly that and which allows to disentangle the momentum modes contributing to the pole-$\MSb$ mass relation and resum the logarithms of quark mass ratios which arise in this multi-scale problem.

For simplicity, in this talk only the case of the top quark being the external heavy quark is discussed. We refer to Ref.~\cite{Hoang:2017btd} for the general case.

\section{Renormalization Group Framework}\label{sec:RGF}
\subsection{Integrating out the Top and R-Evolution}\label{sec:MatchEvo}

Including the bottom and charm quark as massive flavors, the pole-$\MSb$ mass relation for the top quark can be written in the form~\cite{Hoang:2017btd}
\begin{align}\label{eqn:mpoleMSbar}
 \mpole_t -\, \mbar_t ={}& \mbar_t\,\sum_{n=1}^\infty\,a_n(n_t+1)\,\left(\frac{\alpha_s^{(n_t+1)}(\mbar_t)}{4\pi}\right)^n\nonumber\\
 &+ \mbar_t\left[\delbar_t^{\,(t,b,c)}(1,r_{bt},r_{ct}) + \delbar_t^{\,(b,c)}(r_{bt},r_{ct}) + \delbar_t^{\,(c)}(r_{ct})\right]\,,
\end{align}
where $n_t=5$ denotes the number of quark flavors lighter than the top, and the perturbative coefficients $a_n(n_t+1)\equiv a_n(n_t+1,0)$ are now describing only the QCD corrections from gluons and $n_t+1$ massless virtual quark flavors.
The terms $\delbar_t$ contain the mass corrections coming from the top quark on-shell self-energy diagrams with insertions of virtual massive quark loops and can be written in a perturbative expansion in $\alpha_s^{(n_t+1)}$. The superscripts of the form $(q,q^\prime,\dots)$ indicate that each included diagram contains at least one insertion of the massive quark $q$ and in addition all possible insertions of the (lighter) massive quarks $q^\prime,\dots$ as well as of massless quark and gluonic loops. $b$ and $c$ refer to the bottom and charm quark respectively. From each diagram the corresponding diagram with all the quark loops in the massless limit is subtracted. The fraction $r_{qq^\prime} \equiv \mbar_q/\mbar_{q^\prime}$ stands for the ratio of $\MSb$ masses for the quarks $q$ and $q^\prime$.

\begin{figure}
	\center
	\includegraphics[scale=.16]{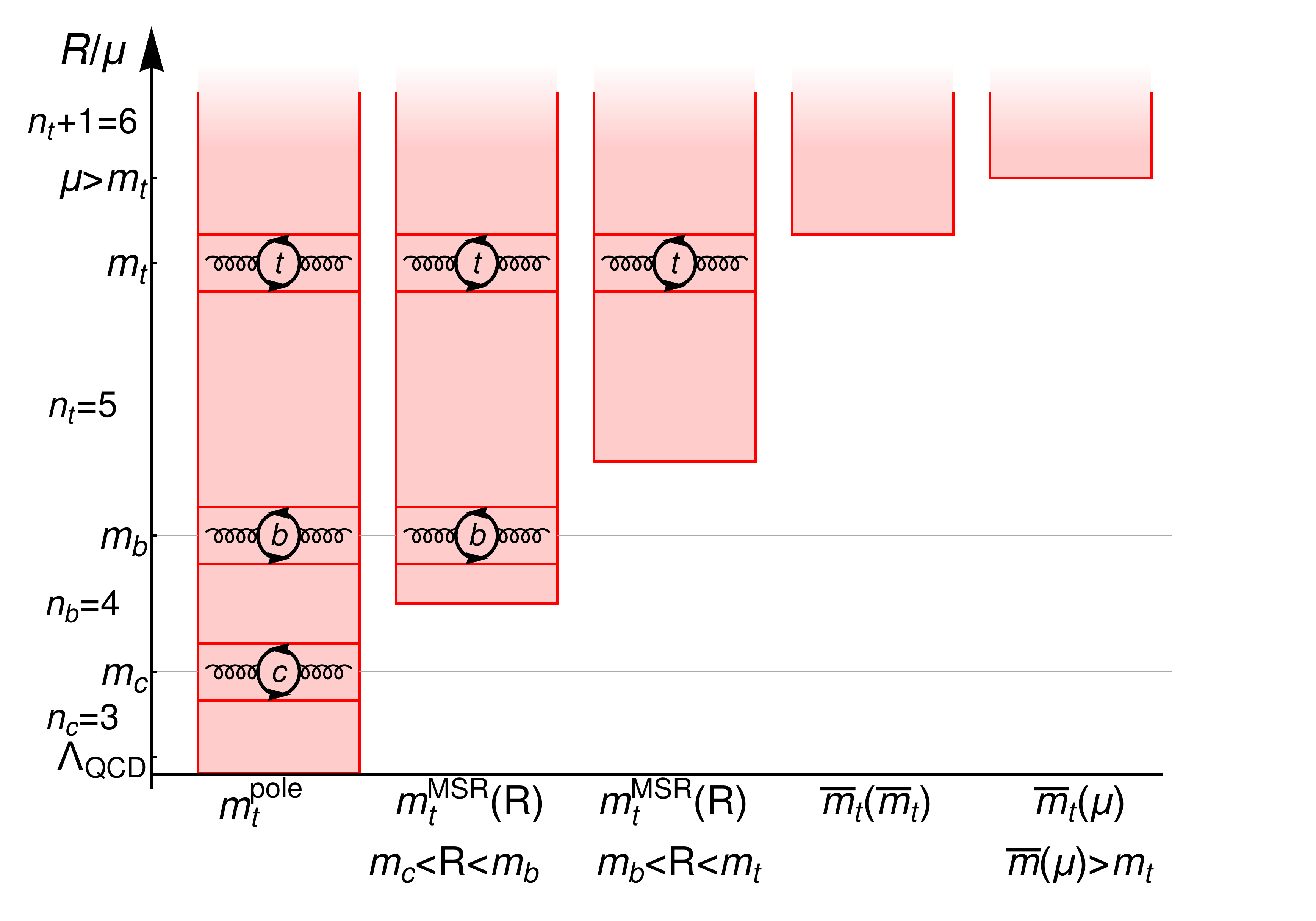}
	\includegraphics[scale=.16]{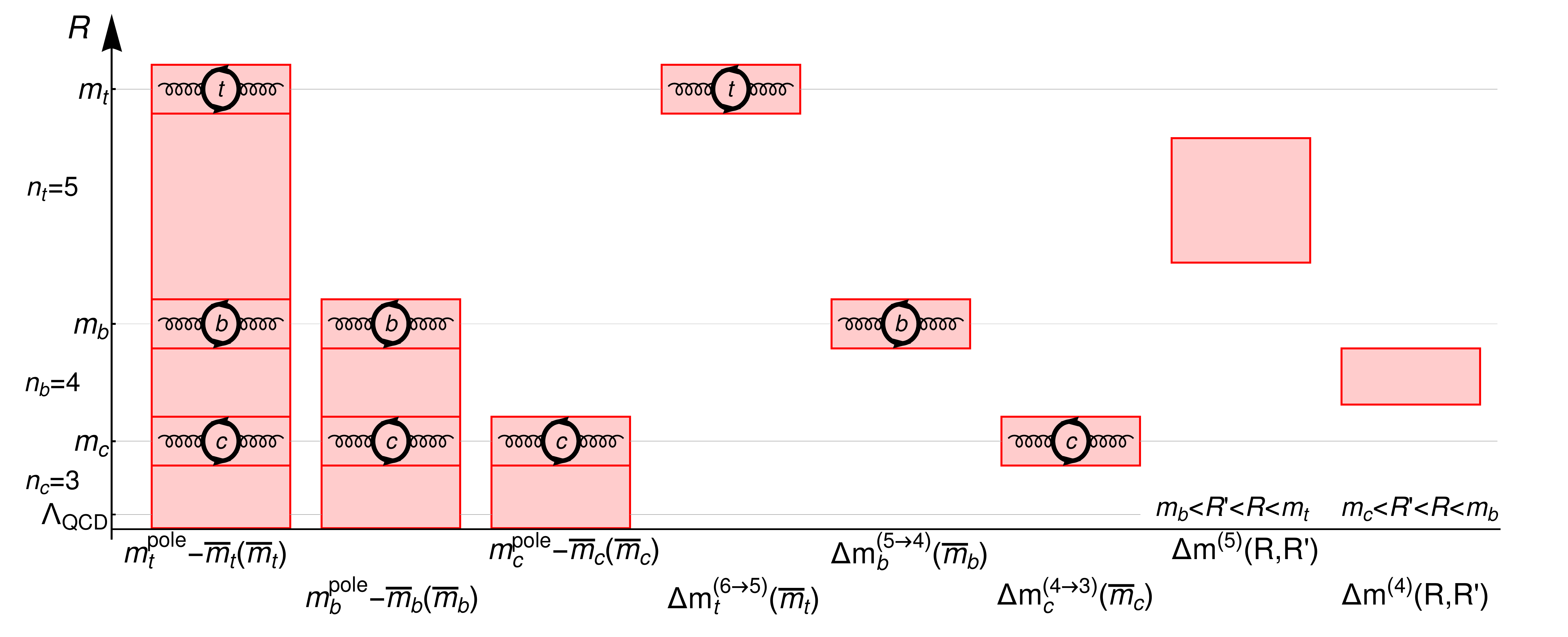}
	\caption{
	\textit{Left:} Graphical illustration of the physical contributions contained in the pole, MSR and $\MSb$ mass schemes coming from the different momentum scales for the case of the top quark. The quark loops stand for the contributions of the virtual massive quark loops contained in the masses.
	\textit{Right:} Graphical illustration for pole-$\MSb$ mass differences, the MSR-$\MSb$ mass matching corrections and MSR mass differences for different $R$ scales. \label{fig:matchrun}\label{fig:massschemes}
	}
\end{figure}

To set up the renormalization group framework and disentangle the different momentum regions below the top mass scale, we use the natural MSR mass $m_t^\MSR(R)$ which was introduced in Ref.~\cite{Hoang:2008yj,Hoang:2017suc}, adapted to account for the effects of massive lighter quarks. In the presence of massive bottom and charm quarks (i.e.\ for $R$ scales between top and bottom mass) the top quark MSR mass is then defined through~\cite{Hoang:2017btd}\footnote{In Ref.~\cite{Mateu:2017hlz} a different version of the MSR mass was suggested where for the lighter quark mass corrections in the second term of Eq.~\eqref{eqn:mpoleMSR} all factors $\mbar_t$ (i.e.\ the overall factor and in the $r_{qt}$ ratios) were also replaced by $R$. The R-evolution of the MSR mass in this scheme depends in a more complicated way on the lighter quark masses: the nontrivial light quark mass corrections fully enter the R-evolution equations and the threshold corrections at the lighter quark mass thresholds are modified. In contrast, in our scheme, the evolution agrees exactly with that for $n_q$ massless quarks for scales just below $m_q$ and the remaining nontrivial light quark mass corrections enter as threshold corrections of the evolution. Thus our scheme adopts the convention that is commonly adopted for the QCD coupling $\alpha_s$ and realizes an analogous separation of evolution and matching/threshold corrections. Both schemes, however, lead to equivalent results in phenomenological applications.}
\begin{align}\label{eqn:mpoleMSR}
 \mpole_t -\msr_t(R) = R\,\sum_{n=1}^\infty\,a_n(n_t)\left(\frac{\alpha^{(n_t)}_s(R)}{4\pi}\right)^n + \mbar_t\,\left[\delta_t^{(b,c)}(r_{bt},r_{ct}) + \delta_t^{(c)}(r_{ct})\right]\,,
\end{align}
where the coefficients $a_n$ are the same as in the pole-$\MSb$ relation Eq.~\eqref{eqn:mpoleMSbar} and $R$ is a momentum scale which is in principle arbitrary, but should be sufficiently larger than $\LQCD$ to stay away from the Landau pole. The terms $\delta_t$ are derived from the respective $\delbar_t$ of Eq.~\eqref{eqn:mpoleMSbar} by (literally) replacing \mbox{$\alpha_s^{(n_t+1)}\to\alpha_s^{(n_t)}$} in the perturbative expansions.
Introducing the MSR mass is useful since the $\MSb$ mass is not adequate to describe scales far below the heavy quark mass scale. The MSR mass achieves two aims in this context: first, the heavy quark is removed as a dynamical degree of freedom from the series (i.e.\ integrated out). Second, the MSR mass introduces linear scale dependence (which is realized in its definition where each factor $\mbar_t$ multiplying the coefficients $a_n$ on the RHS of Eq.~\eqref{eqn:mpoleMSbar} is replaced by the arbitrary momentum scale $R$). Linear scale dependence is crucial in the low momentum region to describe the linear low momentum sensitivity of the pole mass. The replacement of $\mbar_t$ and the removal of the dynamical effects of the top quark in the definition of the MSR mass do not change the asymptotic high order behavior of the series, since the latter is independent of the heavy quark mass and the number of massive flavors. One can interpret the MSR mass as the pole mass minus all self-energy contributions coming from scales below $R$ and all virtual quark mass corrections from quarks lighter than the heavy quark, see the left plot of Fig.~\ref{fig:massschemes}. So the MSR mass represents the proper extension of the $\MSb$ mass concept for renormalization scales below the top quark mass.

Integrating out the top quark leads to a matching coefficient
\begin{equation}
\label{eqn:MSRMSbmatch}
\Delta m_t^{(6\rightarrow 5)}(\mbar_t)=\msr_t(\mbar_t)-\mbar_t,
\end{equation}
which contains the hard corrections coming from the virtual heavy top and therefore does not have any ${\cal O}(\Lambda_{\rm QCD})$ ambiguity, which is also illustrated in the right plot of Fig.~\ref{fig:matchrun}.
$\Delta m_t^{(n_t+1\rightarrow n_t)}(\mbar_t)$ can therefore be computed to high precision using the $a_n$ coefficients which are known up to $\Ord(\alpha_s^4)$~\cite{Tarrach:1980up, Gray:1990yh,Chetyrkin:1999ys,Chetyrkin:1999qi,Melnikov:2000qh,Marquard:2007uj,Marquard:2015qpa,Marquard:2016dcn}, see Tab.~\ref{tab:num}.

The renormalization group equation in $R$ resulting from Eq.\eqref{eqn:mpoleMSR} is linear in $R$ and is called the R-evolution equation. For an arbitrary heavy quark flavor $Q$, it takes the form
\begin{equation}\label{eqn:revolvdef}
 R\frac{\dd}{\dd R}m_Q^{\mathrm{MSR}}(R)=-\,R\,\gamma^{\,R,(n_Q)}(\alpha^{(n_Q)}_s(R))
 =-\,R\sum_{n=0}^\infty\gamma_n^{\,R,(n_Q)}\bigg(\frac{\alpha^{(n_Q)}_s(R)}{4\pi}\bigg)^{n+1}\,,
\end{equation}
where the coefficients $\gamma_n^{\,R,(n_Q)}$ are known up to four loops and given in Refs.~\cite{Hoang:2008yj,Hoang:2017suc}. It is easy to see that Eq.~\eqref{eqn:revolvdef} is renormalon-free since the renormalon ambiguity of the series proportional to $R$ is independent of $R$ and therefore cancels when differentiated.
The solution of the R-evolution equation Eq.~\eqref{eqn:revolvdef} can be used to relate MSR masses at different values of the scale $R$ in a renormalon free way without picking up large logarithms. This solution can be written as
\begin{equation}\label{eqn:rrge}
\Delta m^{(n_Q)}(R,R^\prime) = \msr_Q(R^\prime) - \msr_Q(R) = \sum_{n=0}^\infty \gamma_n^{R,(n_Q)} \int_{R^\prime}^R \dd R\,\left(\frac{\alpha^{(n_Q)}_s(R)}{4\pi}\right)^{n+1}\,,
\end{equation}
where $\Delta m^{(n_Q)}(R,R^\prime)$ represents the self-energy contributions to the mass in the presence of $n_Q$ active dynamical flavors coming from the scales between $R^\prime$ and $R$. This is illustrated in the right plot of Fig.~\ref{fig:matchrun}.

\subsection{Top-Bottom and Bottom-Charm Mass Matching}
\label{sec:tbbcmatch}

As the next step one can, successively, integrate out the lighter massive flavors, which in the case of the top quark applies to the bottom and charm quarks.
 
In a first step one compares the pole-MSR mass relation for the top quark of Eq.~\eqref{eqn:mpoleMSR} to the pole-$\MSb$ mass relation \textit{for the next lighter massive quark}, i.e.\ the bottom
\begin{align}\label{eqn:mpoleMSbarbottom}
 \mpole_b -\, \mbar_b = \mbar_b\,\sum_{n=1}^\infty\,a_n(n_b+1)\,\left(\frac{\alpha_s^{(n_b+1)}(\mbar_b)}{4\pi}\right)^n+\mbar_b\left[\delbar_t^{\,(b,c)}(1,r_{cb}) + \delbar_t^{\,(c)}(r_{cb})\right]\,,
\end{align}
with $n_b+1=n_t=5$. For $R=\mbar_b$ the RHS is identical in the approximation that in the virtual quark loops all $n_t$ lighter quarks (including the bottom quark) are treated as massless. This identity is a consequence of heavy quark symmetry~\cite{Isgur:1989vq} and is also valid when comparing the bottom to the charm quark.

The difference of these two expressions encodes the heavy quark symmetry breaking corrections coming from the finite virtual charm and bottom quark masses and the resulting matching relation reads
\begin{align}\label{eqn:tbmatch}
\delta m_{b,c}^{(t\rightarrow b)}(\mbar_b,\mbar_c)={}&\left[\mpole_t-\msr_t(\mbar_b)\right] - \left[\mpole_b-\mbar_b\right]\nonumber\\
={}&\mbar_t \left[ \delta_t^{(b,c)}(r_{bt},r_{ct}) + \delta_t^{(c)}(r_{ct})\right] - \mbar_b \left[ \overline{\delta}_b^{(b,c)}(1,r_{cb}) + \bar{\delta}_b^{(c)}(r_{cb})\right]\,.
\end{align}

The individual $\delta_n$ terms in the second line of Eq.~\eqref{eqn:tbmatch} carry infrared sensitive contributions and therefore contain an ${\cal O}(\LQCD)$ renormalon ambiguity which leads to very bad perturbative behavior. In Eq.~\eqref{eqn:tbmatch} however,
these renormalon ambiguities mutually cancel such that the top-bottom mass matching $\delta m_{b,c}^{(t\rightarrow b)}(\mbar_b,\mbar_c)$ is a short distance quantity and shows excellent convergence, see Fig.~\ref{fig:HQbreaking}.

After doing the top-bottom mass matching the problem of integrating out the bottom quark is analogous to integrating out the top quark in Sec.~\ref{sec:MatchEvo} which results in another matching contribution $\Delta m_b^{(5\rightarrow 4)}(\mbar_b)=\msr_b(\mbar_b)-\mbar_b$, containing the corrections coming from virtual bottom quarks.

\subsection{Putting the Pieces Together}
\label{sec:together}

Through successive R-evolution with the appropriate flavor number (see Eq.~\eqref{eqn:rrge}) and matching at the mass thresholds of the bottom and charm quark (see Eqs.~\eqref{eqn:MSRMSbmatch}, \eqref{eqn:tbmatch} and their generalizations to the appropriate flavors) we can now decouple the different momentum regions in the pole-$\MSb$ relation of Eq.~\eqref{eqn:mpoleMSbar}.
The resulting formula for the top quark pole mass reads
\begin{align}\label{eqn:mtpolembmc}
 \mpole_t=\mbar_t&+\Delta m_t^{(6\to 5)}(\mbar_t)+\Delta m^{(5)}(\mbar_t,\mbar_b)+\delta m_{b,c}^{(t\to b)}(\mbar_b,\mbar_c)\nonumber\\
 &+\Delta m_b^{(5\to 4)}(\mbar_b)+\Delta m^{(4)}(\mbar_b,\mbar_c)+\delta m_{c}^{(b\to c)}(\mbar_c)\\
&+\Delta m_c^{(4\to 3)}(\mbar_c)+\Delta m^{(3)}(\mbar_c,R)+ R\sum_{n=1}^\infty a_n(\nl=3,0)\left(\frac{\alpha_s^{(3)}(R)}{4\pi}\right)^n\,,\nonumber
\end{align}
where all logarithms $\log(\mbar_b/\mbar_t)$ and $\log(\mbar_c/\mbar_b)$ are systematically resummed. All quantities except for the last term are free from an $\Ord(\LQCD)$ renormalon ambiguity and can be evaluated to high precision using the available 4-loop expressions for $a_n$~\cite{Tarrach:1980up, Gray:1990yh,Chetyrkin:1999ys,Chetyrkin:1999qi,Melnikov:2000qh,Marquard:2007uj,Marquard:2015qpa,Marquard:2016dcn} and 3-loop expressions for the mass corrections~\cite{Gray:1990yh,Bekavac:2007tk}, see Tab.~\ref{tab:num}. The renormalon ambiguity is contained solely in the $R$-dependent last term which is just equal to $m^\pole_c-\msr_c(R)$. This relation specifies the charm quark pole mass ambiguity, and it fully encodes the top and bottom quark pole mass ambiguities due to heavy quark symmetry~\cite{Isgur:1989vq}.
The occurring contributions are illustrated in the right plot of Fig.~\ref{fig:matchrun}.

\begin{table}
\centering
\begin{tabular}{|c|c|c|c|c|}
	\hline
	${\cal O}(\alpha_s^n)$ & $n=1$ & $n=2$ & $n=3$ & $n=4$ \\\hline\hline
	$\Delta m^{(5)}(\mbar_t,\mbar_b)$ &  $8.536 \pm 1.008$ & $9.336 \pm 0.225$ & $9.368 \pm 0.035$ & $9.331 \pm 0.016$\\
	$\Delta m^{(4)}(\mbar_b,\mbar_c)$ &  $0.337 \pm 0.098$ & $0.419 \pm 0.063$ & $0.434 \pm 0.026$ & $0.423 \pm 0.017$\\\hline
	$\Delta m_t^{(6\rightarrow 5)} (\mbar_t)$ & 0 & $0.021\pm 0.004$ & $0.033\pm 0.003$ & $0.032\pm 0.001$\\
	$\Delta m_b^{(5\rightarrow 4)} (\mbar_b)$ & 0 & $0.003 \pm 0.001$ & $0.006 \pm 0.002$ & $0.004 \pm 0.001$\\\hline
	$\delta m_{b,c}^{(t\rightarrow b)} (\mbar_b,\mbar_c)$ & 0 & $0.007\pm 0.004$ & $0.006\pm 0.001$ & -\\
	$\delta m_{c}^{(b\rightarrow c)} (\mbar_c)$ & 0 & $0.004 \pm 0.002$ & $0.004 \pm 0.001$ & -\\\hline
\end{tabular}
\caption{
Numerical values of the universal building blocks of Eq.~\eqref{eqn:mtpolembmc}, all in GeV. For the quark mass values $\mbar_t=163$~GeV, $\mbar_b=4.2$~GeV and $\mbar_c=1.3$~GeV was used.
}\label{tab:num}
\end{table}

\section{Some Applications}\label{sec:applications}

\subsection{Light Virtual Quark Mass Corrections at ${\cal O}(\alpha_s^4)$}
\label{sec:lightvirtual}

\begin{figure}
\center
\includegraphics[width=0.4\textwidth]{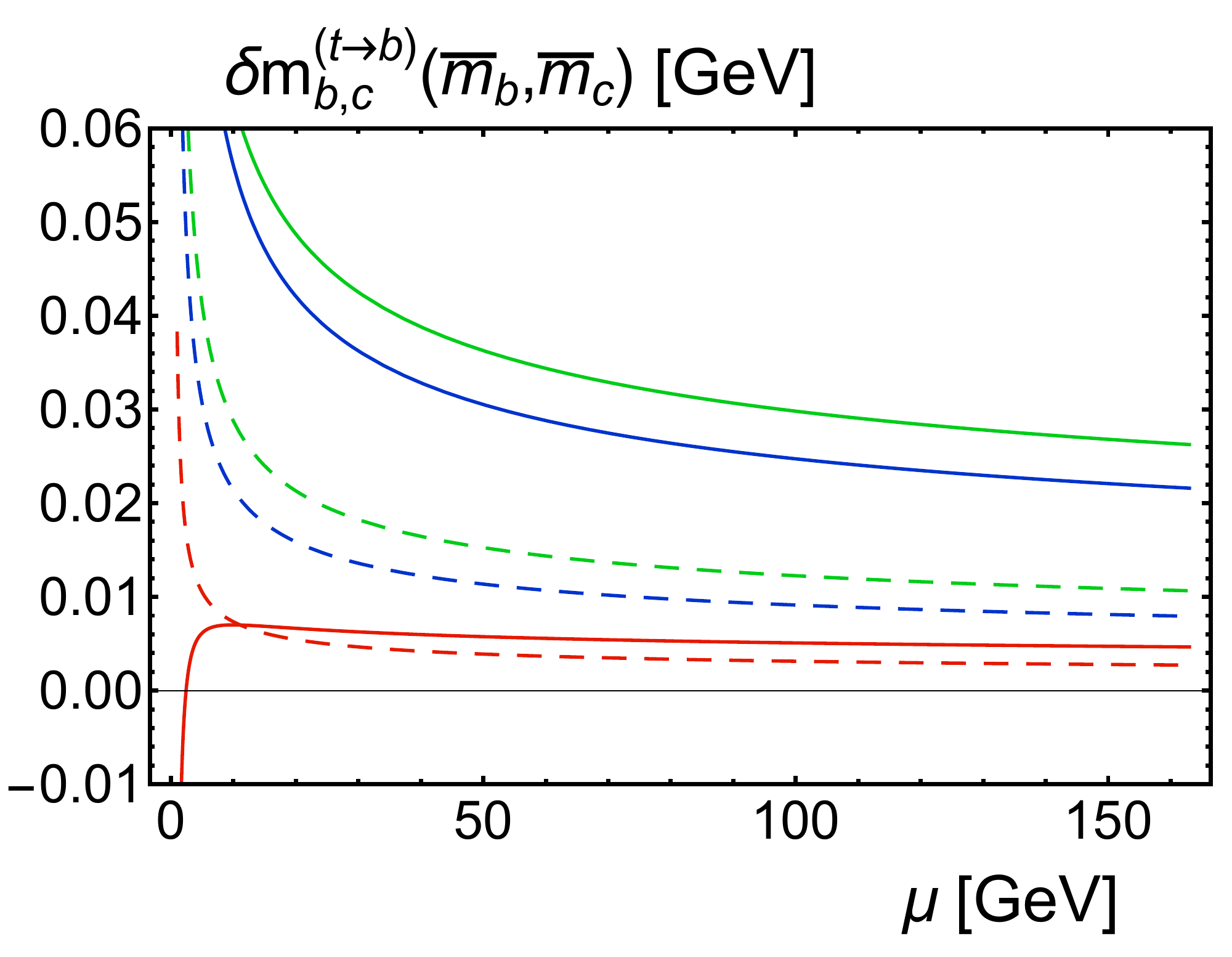}\qquad
\includegraphics[width=0.4\textwidth]{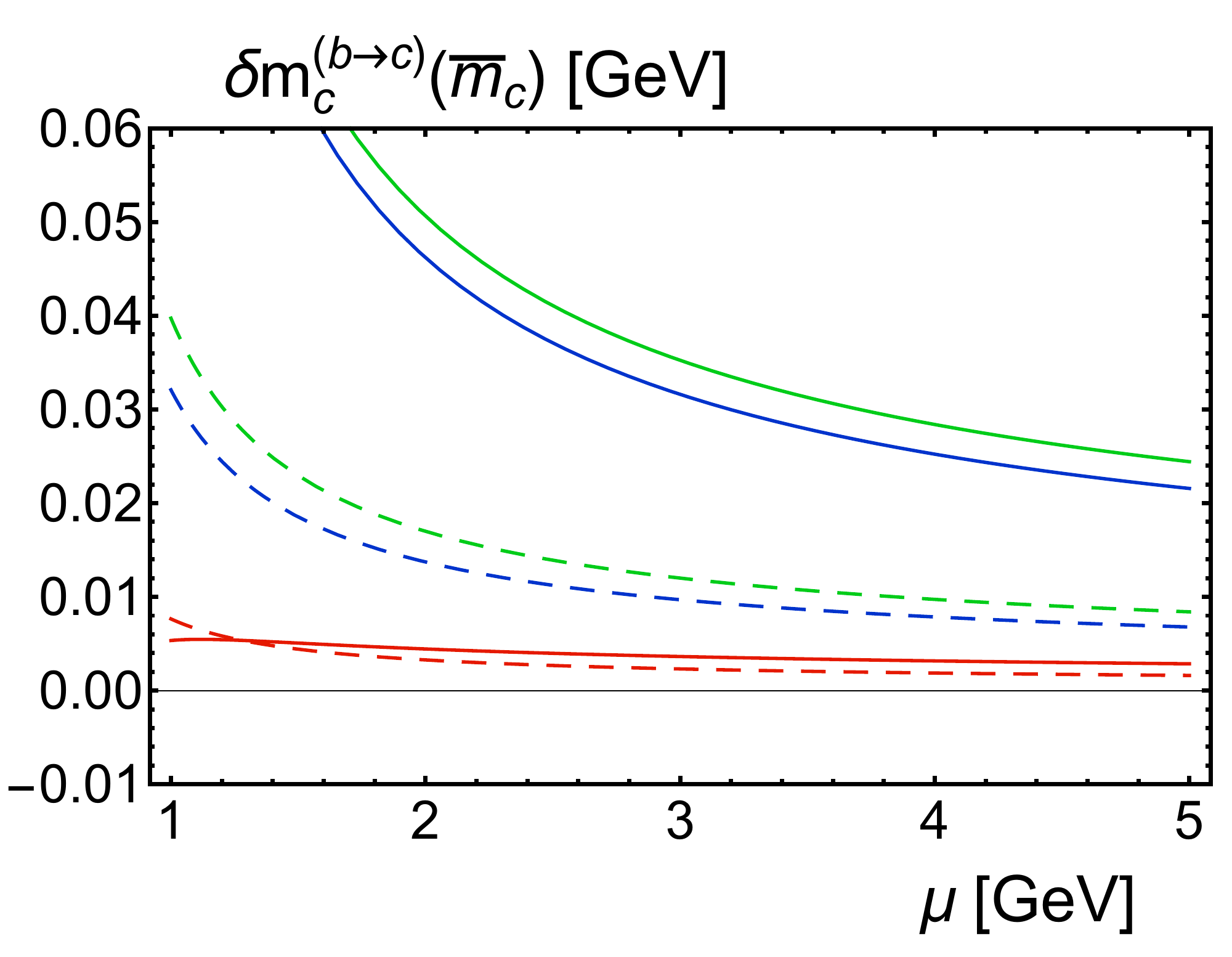}
\caption{\label{fig:HQbreaking} \textit{Left:} Top-MSR bottom-$\MSb$ mass matching correction $\delta m_{b,c}^{(t\rightarrow b)}(\mbar_b,\mbar_c)$ at ${\cal O}(\alpha_s^2)$ (red dashed curve) and ${\cal O}(\alpha_s^3)$ (red solid curve) over the renormalization scale $\mu$, showing excellent perturbative behavior. The virtual bottom and charm mass effects to the top quark self-energy of the first term of the second line in Eq.~\eqref{eqn:tbmatch} (green curves) and the virtual bottom and charm mass effects to the bottom quark self-energy of the second term of the second line of Eq.~\eqref{eqn:tbmatch} (blue curves) at ${\cal O}(\alpha_s^2)$ (dashed) and ${\cal O}(\alpha_s^3)$ (solid) individually showing very bad perturbative behavior. For the masses of the top, bottom and charm quarks the values ($\mbar_t,\mbar_b,\mbar_c$) = ($163,4.2,1.3$)~GeV are used. \textit{Right:} The bottom-MSR charm-$\MSb$ mass matching correction $\delta m_{c}^{(b\rightarrow c)}(\mbar_c)$ at ${\cal O}(\alpha_s^2)$ (red dashed curve) and ${\cal O}(\alpha_s^3)$ (red solid curve) over the renormalization scale $\mu$. The virtual charm mass effects to the bottom quark self-energy (green curves) and the virtual charm mass effects to the charm quark self-energy (blue curves) are shown at ${\cal O}(\alpha_s^2)$ (dashed) and ${\cal O}(\alpha_s^3)$ (solid).
}
\end{figure}

As mentioned in Sec.~\ref{sec:tbbcmatch}, the mass matching contributions show excellent convergence although their individual contributions from the mass corrections in the second line of Eq.~\eqref{eqn:tbmatch} bear very large renormalon contributions.
In the left plot of Fig.~\ref{fig:HQbreaking} the top-MSR bottom-$\MSb$ mass matching correction $\delta m_{b,c}^{(t\rightarrow b)}(\mbar_b,\mbar_c)$ of Eq.~\eqref{eqn:tbmatch} is displayed as a function of the renormalization scale $\mu$ at ${\cal O}(\alpha_s^2)$ (red dashed line) and ${\cal O}(\alpha_s^3)$ (red solid line) for ($\mbar_t,\mbar_b,\mbar_c$) = ($163,4.2,1.3$)~GeV. The matching correction at ${\cal O}(\alpha_s^3)$ amounts to $6$~MeV and has a scale variation of only $1$~MeV for $\mbar_b\leq\mu\leq\mbar_t$. Compared to the ${\cal O}(\alpha_s^2)$ result we see a strong reduction of scale dependence. The plot also shows the virtual bottom and charm mass effects in the top quark self-energy (green curves) and the virtual bottom and charm mass effects to the bottom quark self-energy (blue curves), i.e.\ the first and second term of the second line of Eq.~\eqref{eqn:tbmatch}, at ${\cal O}(\alpha_s^2)$ (dashed) and ${\cal O}(\alpha_s^3)$ (solid). Both types of contributions each are quite large and furthermore do not at all converge. The ${\cal O}(\alpha_s^3)$ corrections are even bigger than the ${\cal O}(\alpha_s^2)$ corrections, which indicates that the corresponding asymptotic large order behavior already dominates the ${\cal O}(\alpha_s^2)$ and ${\cal O}(\alpha_s^3)$ corrections. An analog plot for the matching contribution $\delta m_{c}^{(b\rightarrow c)}(\mbar_c)$, describing the heavy quark symmetry breaking corrections for the bottom quark coming from the finite charm quark mass can be seen on the right side of Fig.~\ref{fig:HQbreaking}.

This cancellation is expected theoretically due to heavy quark symmetry~\cite{Isgur:1989vq}. However, the facts that the overall size of the matching corrections only amounts to a few MeV, and that the corrections are only around $1$~MeV already at ${\cal O}(\alpha_s^3)$ allows to draw interesting conceptual implications for the large order asymptotic behavior of the virtual quark mass corrections in the mass relations of Eq.~\eqref{eqn:mpoleMSbar} because we can expect the $\Ord(\alpha_s^4)$ matching corrections amount to less than $1$~MeV. As a consequence we can predict the yet uncalculated virtual quark mass corrections at ${\cal O}(\alpha_s^4)$ to within a few percent without an additional loop calculation by approximating the ${\cal O}(\alpha_s^4)$ correction in the mass matching by zero.

Let's consider the matching correction $\delta m_{q}^{(Q\rightarrow q)} (\mbar_q)$ between the MSR mass of heavy quark $Q$ and the $\MSb$ mass of the next lighter massive quark $q$ assuming the massless approximation for all quarks lighter than quark $q$, i.e.\ $\nq = n_q+1 = \nl+1$ and $n_\ell=n_q$ being the number of massless quarks. This situation applies to the matching relation for the top-MSR and the bottom $\MSb$ masses for a massless charm quark or to the matching relation between the bottom-MSR and the charm-$\MSb$ masses.

For $\mu=\mbar_Q$, we can provide the very simple closed analytic expression
\begin{align}\label{eqn:delta4predict2}
 \delta_{Q,4}^{(q)}(r_{qQ}) \approx r_{qQ}\bigg[\,\delta_{q,4}^{(q)}(1) - \left(6\,\betazero\delta_{q,3}^{(q)}(1) + 4\,\betaone\delta_2(1) \right)\ln\left(r_{qQ}\right) + 12\,\delta_2(1)\left(\betazero\ln(r_{qQ})\right)^2\bigg]\,.
\end{align}
The coefficients $\delta_{q,n}^{(q)}(1)$ describe the corrections from virtual massive loops of the heavy quark $q$ to the $q$ self-energy and are known up to $\Ord(\alpha_s^4)$ from the full $a_n$ coefficients computed in Refs.~\cite{Tarrach:1980up, Gray:1990yh,Chetyrkin:1999ys,Chetyrkin:1999qi,Melnikov:2000qh,Marquard:2007uj,Marquard:2015qpa,Marquard:2016dcn}.

\begin{figure}
\center
 \includegraphics[width=.4\textwidth]{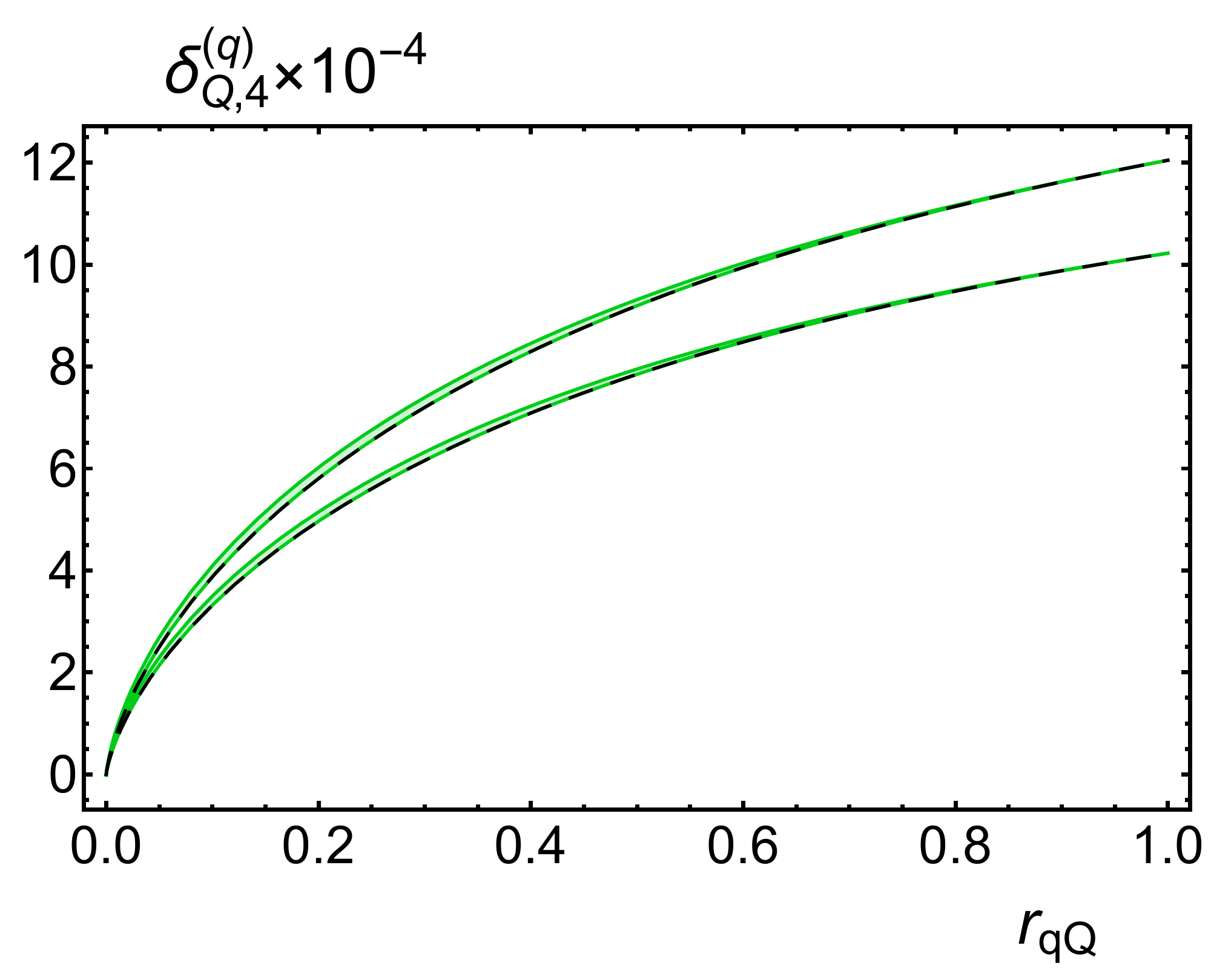}\qquad
 \includegraphics[width=.4\textwidth]{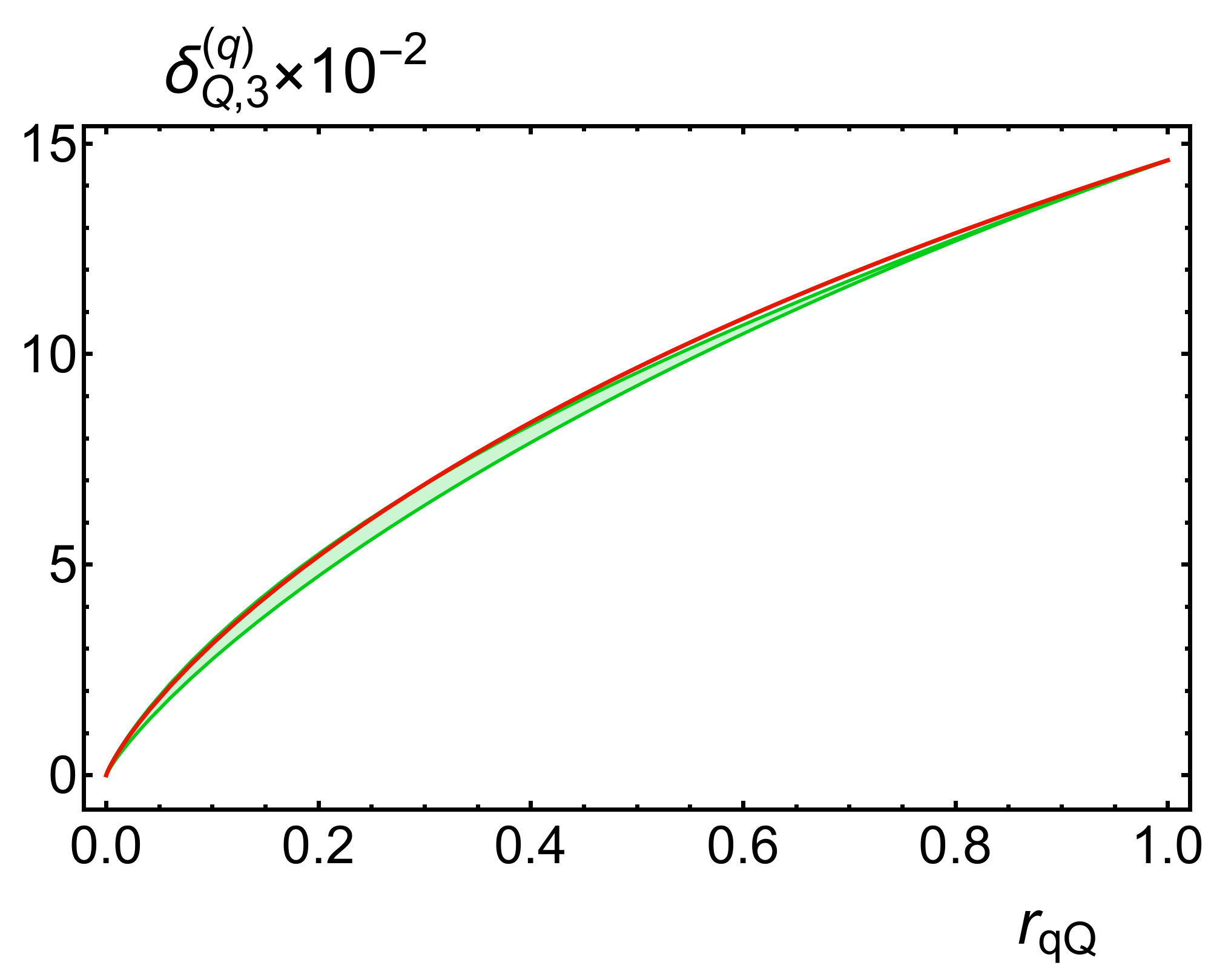}
 \caption{\label{fig:deltapredict}\label{fig:deltapredicta}\label{fig:deltapredictb}
 \textit{Left:} Prediction for the ${\cal O}(\alpha_s^4)$ virtual quark mass correction $\delta_{Q,4}^{(q)}(r_{qQ})$ for $\mbar_q\leq\mu\leq\mbar_Q$ (green bands) for $\nq=\nl+1=5$ (lower band) and $\nq=\nl+1=4$ (upper band). The black dashed lines show the prediction for $\mu=\mbar_Q$. \textit{Right}: The ${\cal O}(\alpha_s^3)$ virtual quark mass correction $\delta_{Q,3}^{(q)}(r_{qQ})$ for $\nq=\nl+1=5$ (red curve). The green band is the prediction for $\delta_{Q,3}^{(q)}(r_{qQ})$ using the method used in the left plot showing excellent agreement to the exact result within errors. 
 }
\end{figure}

In the left plot of Fig.~\ref{fig:deltapredicta} we show the prediction of $\delta_{Q,4}^{(q)}(r_{qQ})$ for the top ($n_Q=n_t=5$, lower band) and bottom ($n_Q=n_b=4$, upper band) with a scale variation of $\mbar_b\leq\mu\leq\mbar_t$ and $\mbar_c\leq\mu\leq\mbar_b$ respectively. The curves for Eq.~\eqref{eqn:delta4predict2} are shown as the black dashed lines. The uncertainty amounts to $\pm3\%$ (for $r_{qQ}\lesssim0.1$) or smaller (for $r_{qQ}>0.1$). The reliability of the prediction and uncertainty estimate was additionally tested by ``predicting'' the already exactly known $\delta_{Q,3}^{(q)}(r_{qQ})$ (red curve) using the same method, see the right plot of Fig.~\ref{fig:deltapredictb}. The prediction is fully compatible with the exact result and the uncertainty amounts to $\pm10\%$ (for $r_{qQ}\lesssim0.1$) or smaller (for $r_{qQ}>0.1$).

This method can be generalized to arbitrary high orders in $\alpha_s$ by using the known asymptotic behavior of the coefficients $a_n$, as well as to the case of having a larger number of lighter massive quarks.

\subsection{Pole Mass Differences}
\label{sec:polediff}

Due to heavy quark symmetry, the difference of two heavy quark pole masses is free of ${\cal O}(\LQCD)$ renormalon ambiguities and can determined to high precision. The matching and R-evolution of the MSR mass allow us to systematically sum logarithms of the mass ratios that would remain unsummed in a fixed-order calculation~\cite{Hoang:2017suc}.
The resulting relations between the top, bottom and charm quark pole masses read
\begin{align}
 \mpole_t - \mpole_b &= \left[ \mbar_t - \mbar_b \right] + \Delta m_t^{(6\rightarrow 5)}(\mbar_t) + \Delta m^{(5)}(\mbar_t,\mbar_b) + \delta m_{b,c}^{(t\rightarrow b)}(\mbar_b,\mbar_c) \label{eqn:tbpole} \,,\\
 \mpole_b - \mpole_c &= \left[ \mbar_b - \mbar_c \right] + \Delta m_b^{(5\rightarrow 4)}(\mbar_b) + \Delta m^{(4)}(\mbar_b,\mbar_c) + \delta m_{c}^{(b\rightarrow c)}(\mbar_c) \label{eqn:bcpole} \,,\\
 \mpole_t - \mpole_c &= \left[ \mbar_t - \mbar_c \right] + \Delta m_t^{(6\rightarrow 5)}(\mbar_t) + \Delta m^{(5)}(\mbar_t,\mbar_b) + \delta m_{b,c}^{(t\rightarrow b)}(\mbar_b,\mbar_c) \nonumber\\
 &\hspace{5em} + \Delta m_b^{(5\rightarrow 4)}(\mbar_b) + \Delta m^{(4)}(\mbar_b,\mbar_c) + \delta m_{c}^{(b\rightarrow c)}(\mbar_c) \label{eqn:tcpole} \,.
\end{align}
Each of the mass differences is a sum of the universal matching and evolution building blocks discussed in Sec.~\ref{sec:RGF} which each can be computed to high precision. The numerical evaluation using the values given in Tab.~\ref{tab:num} for the case ($\mbar_t,\mbar_b,\mbar_c$) = ($163,4.2,1.3$)~GeV gives
\begin{align}
 \mpole_t - \mpole_b &= 168.169\pm0.016 \;{\rm GeV} \,, \label{eqn:tbpole2} \\
 \mpole_b - \mpole_c &= \hphantom{16}3.331\pm0.017 \;{\rm GeV} \,,  \label{eqn:bcpole2}  \\
 \mpole_t - \mpole_c &= 171.500 \pm 0.024 \;{\rm GeV}  \,. \label{eqn:tcpole2}
\end{align}
The uncertainties should be considered as conservative estimates of the theoretical uncertainties due to missing higher order corrections.

\subsection{Pole Mass Ambiguity}
\label{sec:ambiguity}

\begin{figure}
\centering
\includegraphics[width=.39\textwidth]{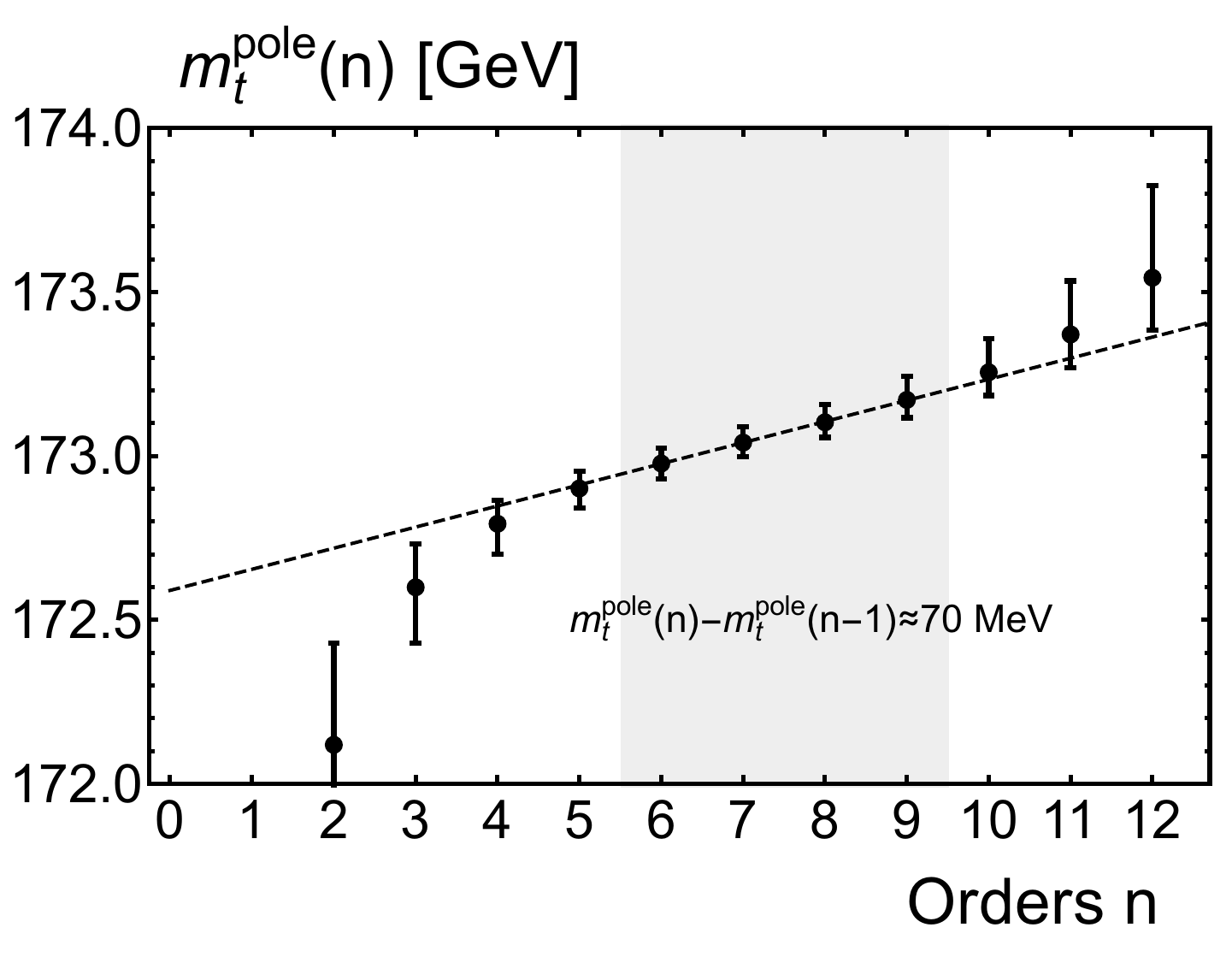}\qquad
\includegraphics[width=.39\textwidth]{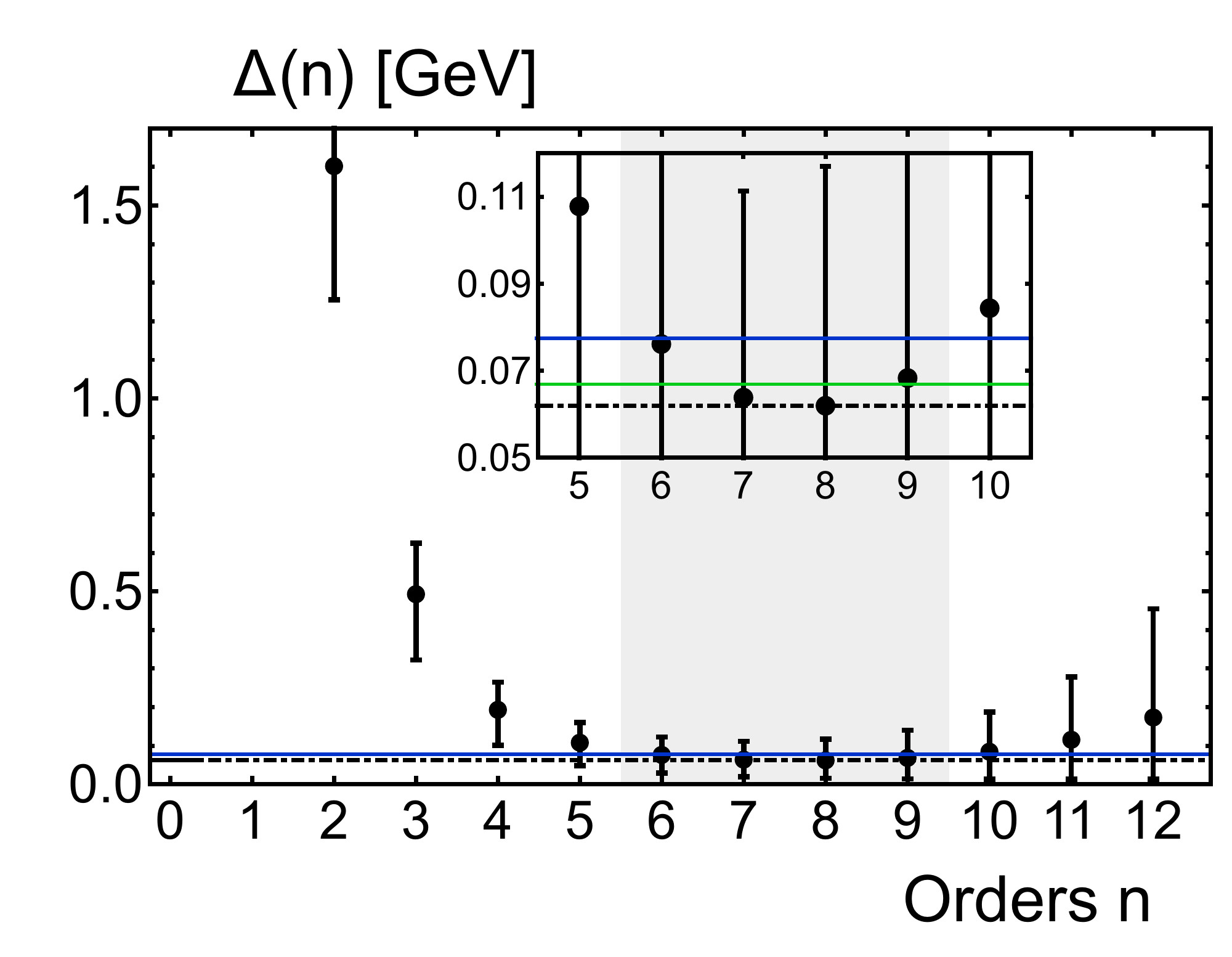}
\raisebox{50pt}{\includegraphics[width=.15\textwidth]{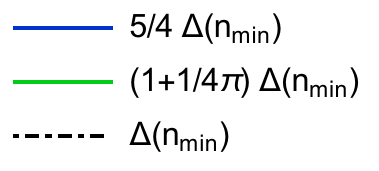}}
\caption{\textit{Left}: Top quark pole mass as a function of order obtained from the MSR mass $\msr_t(\mbar_t)$, $\mbar_t = 163$~GeV for massless bottom and charm quarks. The central dots refer to the renormalization scale $\mu = \mbar_t$ for the strong coupling. The error bars arise from renormalization scale variation $\mbar_t/2 \leq \mu \leq 2\mbar_t$. The shaded stripe represents the region where the series grows almost linearly. \textit{Right:} Size of the corrections $\Delta(n)$ of $\Ord(\alpha_s^n)$ including scale variations from the left panel. The gray stripe represents the region where the corrections are very close to the minimal correction $\Delta(n)$.}\label{fig:ambi1}
\end{figure}

It is well known that for asymptotic series the best possible estimate of the related quantity is obtained when truncating the series at the smallest correction term $\Delta(n_\mathrm{min})$, where $\Delta(n)\equiv m_t^\pole(n)-m_t^\pole(n-1)$, and $m_t^\pole(n)$ is the partial sum at $\Ord(\alpha_s^n)$ of the series for the top quark pole mass that contains the $\Ord(\LQCD)$ pole mass renormalon. The question is how large the uncertainty of this best estimate is in the case of the series describing the relation between the pole and short distance masses. For a truncated converging series the uncertainty is usually estimated by scale variation or by the size of the correction term where the series is truncated. However, in the case of the considered diverging asymptotic series scale variation alone is not useful for error estimation and there is no unique minimal correction term $\Delta(n_\mathrm{min})$ to truncate the series since there is a region in orders $n$ where the terms are almost of the same size~\cite{Bigi:1994em,Beneke:1994sw} (referred to as the ``flat region'') and where $m_t^\pole(n)$ increases linearly with the order. Both can be seen in Fig.~\ref{fig:ambi1} for the series for $m_t^\pole$ obtained from $m_t^\MSR(\mbar_t)$ (i.e.\ $R=\mbar_t$).

Our method to determine the inevitable uncertainty of the series, its ambiguity, takes all this into account and is as follows:
\begin{itemize}
\item We identify the size of the smallest correction term $\Delta(n_\mathrm{min})$ and the range in orders $n$ of numerically close ones $\{n\}_f\equiv\{n:\,\Delta(n)\leq f\,\Delta(n_{\rm min})\}$, where $f\gtrsim1$.
\item We use half of the range of values covered by this region and include renormalization scale variation in a given range as an estimate for the ambiguity. The midpoint of the covered range is taken as the central value.
\end{itemize}
For $f$ one should use a value close to one, but sufficiently large such that the orders where the corrections $\Delta(n)$ are close to $\Delta(n_\mathrm{min})$ (in comparison to the $\Delta(n)$ outside the flat region) are covered. We picked $f=1.25$ to be definite and checked that the outcome is equivalent for variations $6/5\leq f\leq 4/3$, see Fig.~\ref{fig:ambi1}.

\begin{figure}
\centering
\includegraphics[width=.3\textwidth]{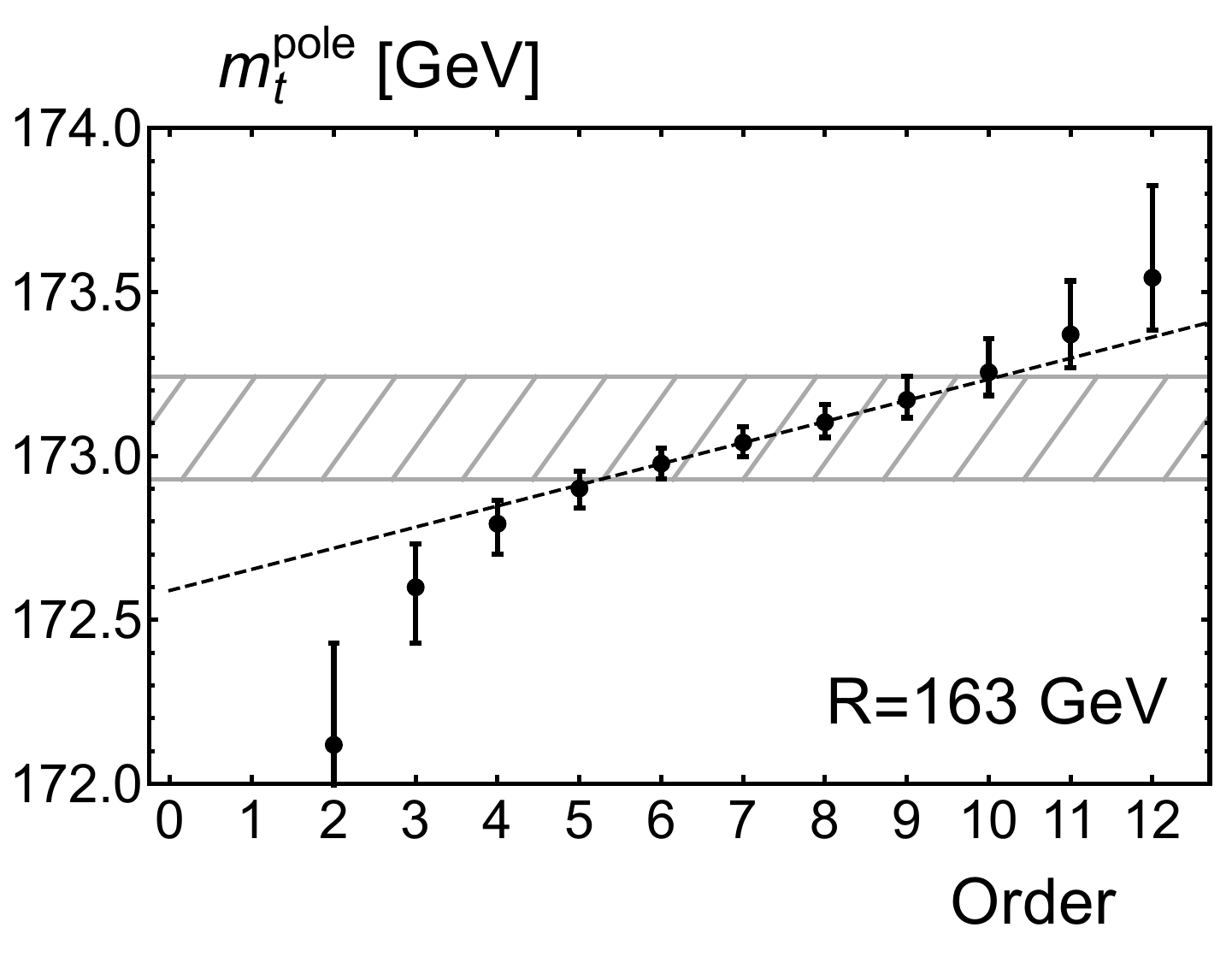}\quad
\includegraphics[width=.3\textwidth]{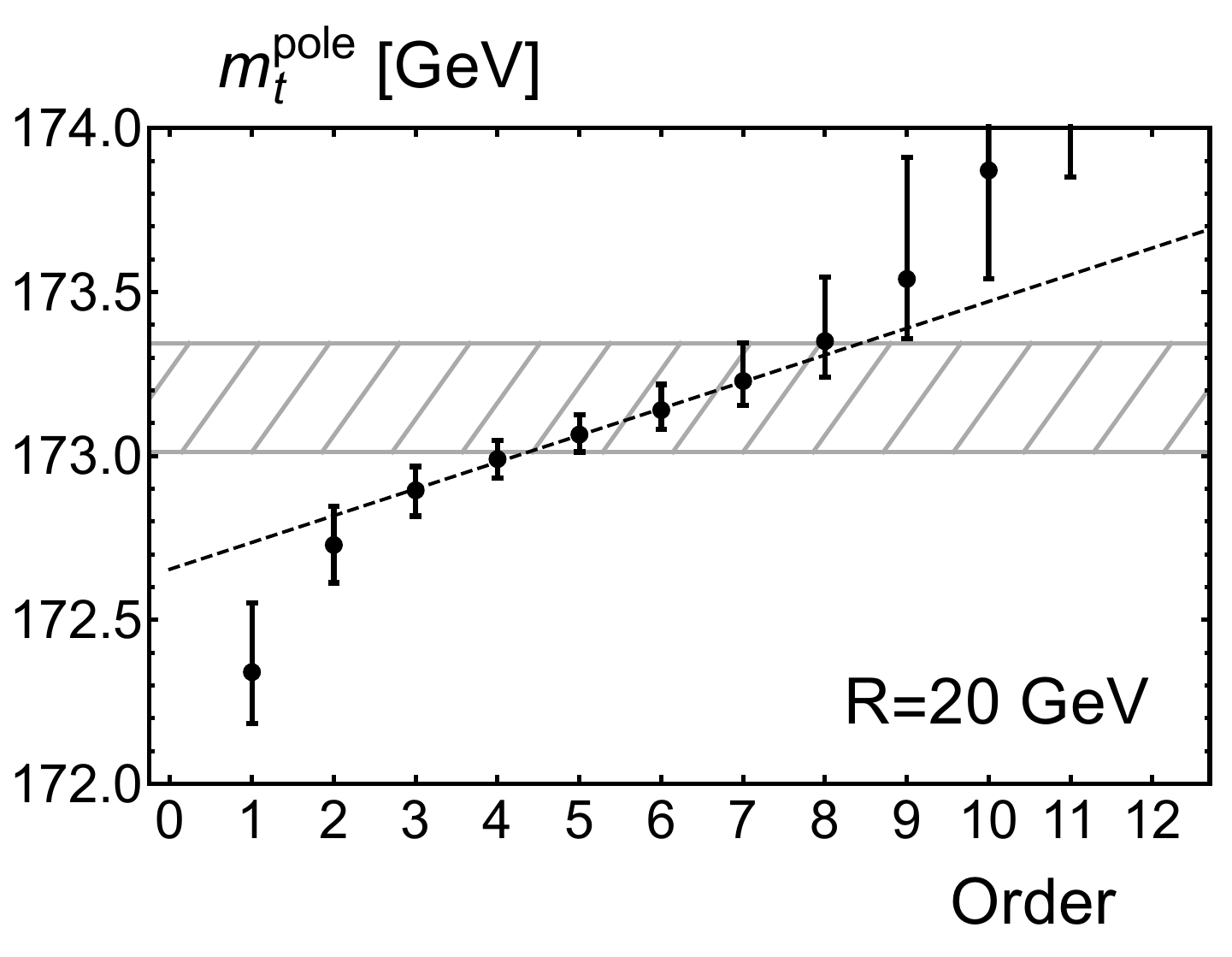}\quad
\includegraphics[width=.3\textwidth]{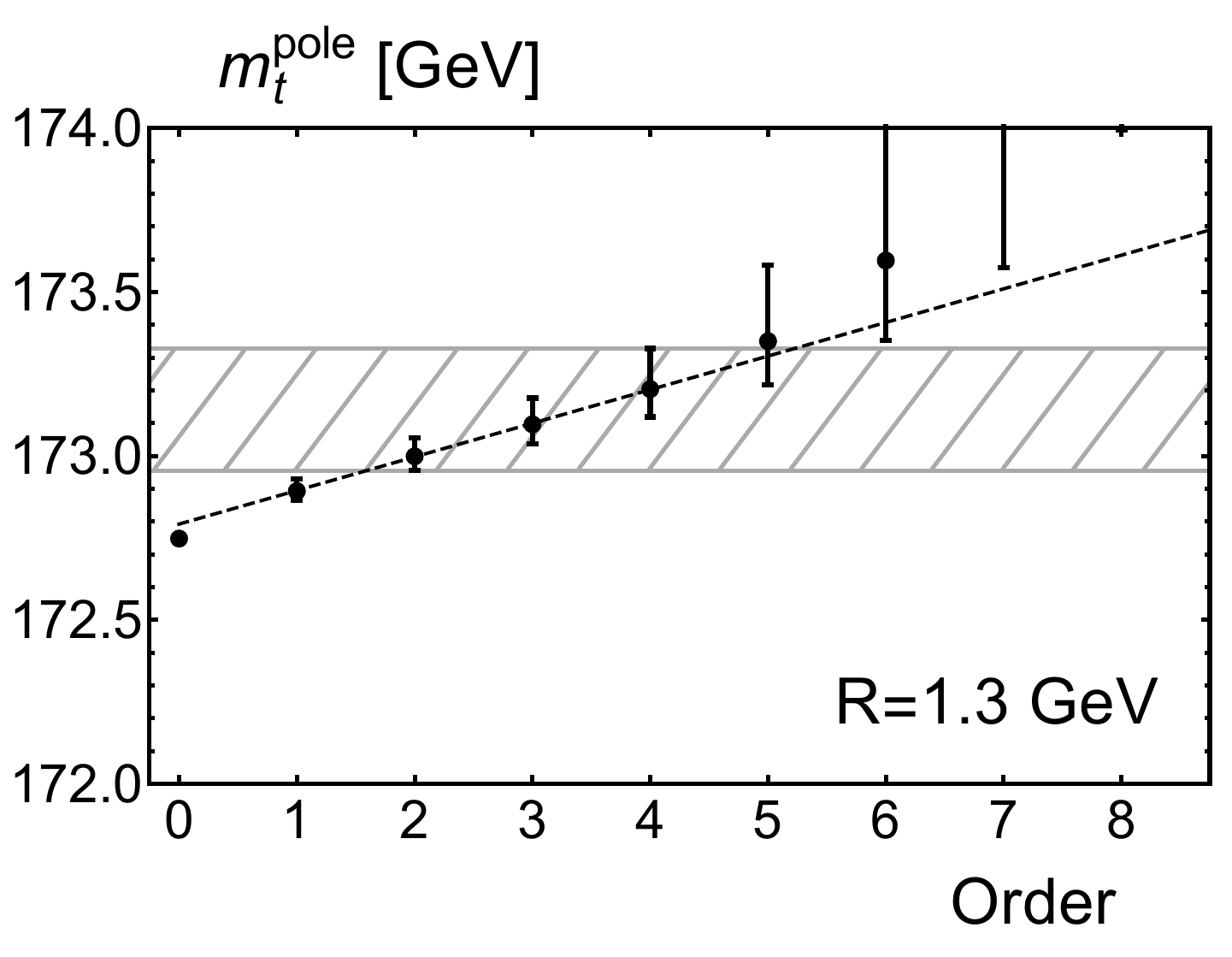}
\caption{Top quark pole mass as a function of order with massless bottom and charm quarks for different values of $R$. The hatched horizontal bands show the best possible estimate as obtained with the method described in Sec.~\ref{sec:ambiguity}.}\label{fig:ambi2}
\end{figure}

Using Eq.~\eqref{eqn:mtpolembmc}, we can cross check that this method is consistent with heavy quark symmetry by varying $R$, since the last term which contains the renormalon is equivalent to the pole-$\MSb$ mass difference of a quark of mass $R$. In Fig.~\ref{fig:ambi2} the top quark pole mass as a function of order obtained from the MSR mass can be seen for different values of $R$, where for illustration the bottom and charm quarks are taken to be massless. Although the minimal correction term varies between about $60$~MeV and $100$~MeV for $R=163$~GeV and $R=1.3$~GeV respectively, the size of the hatched region which represents the ambiguity does hardly change.

Using Eq.~\eqref{eqn:mtpolembmc} it is straightforward to implement bottom and charm quark masses. As a result we obtain an ambiguity of $180$~MeV when bottom and charm quarks are treated massless, and $220$~MeV in the case of a massless charm quark. We obtain an ambiguity of $250$~MeV for the physical case of finite bottom and charm quarks. For more details we refer to Ref.~\cite{Hoang:2017btd}.

The recent estimate of the ambiguity in Ref.~\cite{Beneke:2016cbu} following the prescription of~\cite{Beneke:1998ui} is about $60\%$ smaller and coincides with the size of the minimal correction term $\Delta(n_\mathrm{min})$ and also the scale variation at $\Ord(\alpha_s^\mathrm{min})$ for $R=163$~GeV. For smaller values of $R$ their result for the ambiguity is smaller than $\Delta(n_\mathrm{min})$ and the scale variation at $\Ord(\alpha_s^{n_\mathrm{min}})$.

\section{Comment on arXiv:1712.02796}
\label{sec:comments}

In Ref.~\cite{Nason:2017cxd} one of the authors of Ref.~\cite{Beneke:2016cbu} criticized four aspects of our ambiguity analysis. (a) It is stated that the values of the parameter $f$ ($1.20\leq f\leq 1.33$) were too large for no reason compared to the value $f=1.08$ he claims to be more appropriate. We reply that the value of $f=1.08$ is designed to achieve an ambiguity estimate equal to theirs. Our range of $f$ represents an independent choice motivated by a conservative view on which terms in the flat region (see shaded region in Fig.~\ref{fig:ambi1}) should be taken into account. Since the outcome of the analysis has a rather strong dependence on $f$ in this range, we consider our more conservative choice more appropriate. (b) It is stated that using values of $R$ much below $163$~GeV is unjustified. We reply that our method does not have a significant dependence on $R$ and, furthermore, that the use of $R$ values sufficiently larger than $\LQCD$ is in accordance with heavy quark symmetry. In fact, a consistent argumentation concerning the size of the pole mass ambiguity must yield results that are robust with respect to smaller values of $R$. (c) It is stated that  the $\Ord(\alpha_s^n)$ loop corrections $\Delta(n)$ should not be analyzed at integer values $n$, but as a continuous function of $n$. We reply that we refrain from using a method that relies on using loop corrections continued to arbitrary real values of $n$, because such a treatment is far away from the usual way of dealing with perturbation theory. (d) It is stated that our way to treat the renormalization scale dependence in our ambiguity estimate may be inconsistent because it leads to larger scale variation for smaller $R$ values. We reply that our final result for the ambiguity does not depend on this issue. Furthermore we reply that such behavior is natural for any usual treatment of perturbative series in QCD and that we did not want to rely on a method that is designed to eliminate it on purpose. Overall, we state that the approach employed in Ref.~\cite{Beneke:2016cbu} is in our opinion quite optimistic and that the prescription of Ref.~\cite{Beneke:1998ui} has been adopted without scrutinizing. Our result represents a more conservative treatment that uses heavy quark symmetry as the guiding principle.

\section{Conclusions}
\label{sec:conclusions}

We have provided a renormalization group framework which allows to study the mass effects of virtual massive quark loops in the relation between the pole mass $\mpole_Q$ and short-distance masses such as the $\MSb$ mass $\mbar_Q(\mu)$ of a heavy quark $Q$, where we mean virtual loop insertions of quarks $q$ with $\LQCD < m_q < m_Q$. In this context it is well-known that the virtual loops of a massive quark act as an infrared cut-off on the virtuality of the gluon exchange that eliminates the effects of that quark from the large order asymptotic behavior of the series.

It was examined (i) how the logarithms of mass ratios that arise in this multi-scale problem can be systematically summed to all orders, (ii) the large order asymptotic behavior and structure of the mass corrections themselves and (iii) the consequences of heavy quark symmetry.

Within this framework, we find that the bulk of the lighter virtual quark mass corrections is determined by their large order asymptotic behavior already at ${\mathcal O(\alpha_s^3)}$, which confirms earlier observations made in Refs.~\cite{Hoang:1999us,Hoang:2000fm} and~\cite{Ayala:2014yxa}. We used this property to predict the previously unknown ${\mathcal O(\alpha_s^4)}$ lighter virtual quark mass corrections to within a few percent without an additional loop computation. Furthermore we calculated the differences of the top, bottom and charm quark pole masses with a precision of around $20$~MeV and determined the renormalon ambiguity of the top quark pole mass which amounts to $250$~MeV.

\section*{Acknowledgments}
We  acknowledge partial support by the FWF Austrian Science Fund under the Doctoral Program No.\ W1252-N27 and the Project No.\ P28535-N27.

\bibliographystyle{JHEP}
\bibliography{./sources}

\end{document}